\pdfoutput=1
\documentclass[twocolumn,
               showpacs,
               preprintnumbers,
               nofootinbib,
               prd,
               superscriptaddress,
               10pt,
               notitlepage,
               aps]{revtex4-2}
               
\usepackage{graphicx,amssymb,amsmath,amsthm,amsfonts,epsfig,mathtools}

\usepackage[utf8]{inputenc}
\usepackage{graphicx}
\usepackage{dcolumn}
\usepackage{bm}
\usepackage{amsmath}
\usepackage{color}
\usepackage{soul}
\usepackage[dvipsnames]{xcolor}
\usepackage{hyperref}
\hypersetup{colorlinks=true, citecolor=MidnightBlue,linkcolor=CornflowerBlue, urlcolor=CornflowerBlue, linktocpage=true}
\usepackage{xfrac}
\usepackage{siunitx}

\newcommand{\beqn}{\begin{eqnarray}}
\newcommand{\eeqn}{\end{eqnarray}}
\newcommand{\beq}{\begin{equation}}
\newcommand{\eeq}{\end{equation}}

\def\Phib{\overline{\Phi}}

\def\gbar{\bar{g}}

\def\cala{{\cal A}}

\begin{document}

\title{Treatments and placebos for the pathologies of effective field theories
}

\begin{abstract}
We demonstrate some shortcomings of ``fixing the equations,'' an increasingly popular remedy for time evolution problems of effective field theories (EFTs). We compare the EFTs and their ``fixed'' versions to the UV theories from which they can be derived in two cases: K-essence and nonlinear Proca theory. We find that when an EFT breaks down due to loss of hyperbolicity, fixing does not approximate the UV theory well if the UV solution does not quickly settle down to vacuum. We argue that this can be related to the EFT approximation itself becoming invalid, which cannot be rectified by fixing.
\end{abstract}

\author{Andrew Coates}
\email{andrew.coates.grav@gmail.com}

\author{Fethi M. Ramazano\u{g}lu}
\email{framazanoglu@ku.edu.tr}
\affiliation{Department of Physics, Ko\c{c} University, \\
Rumelifeneri Yolu, 34450 Sariyer, Istanbul, Turkey}

\date{\today}
\maketitle

\section{Introduction}
Even though vector fields are ubiquitous in physics, their self-interaction was only recently shown to be highly problematic. This was originally discovered in the context of cosmology~\cite{Esposito-Farese:2009wbc}, and more recently for strong gravity~\cite{Garcia-Saenz:2021uyv,Silva:2021jya,Demirboga:2021nrc}. A particularly interesting case is the dynamical breakdown of the nonlinear generalization of the Proca theory, for which initially healthy configurations naturally evolve to a point where time evolution cannot be continued~\cite{Clough:2022ygm,Mou:2022hqb,Coates:2022qia}.\footnote{Coordinate singularities can be easily mistaken for the physical breakdown. See \textcite{Coates:2022nif,Coates:2023dmz} for the subtleties of this issue.}  This is particularly unnerving since such vector fields find a wide range of applications in many areas of physics~\cite{Esposito-Farese:2009wbc,ATLAS:2017fur,DeFelice:2016yws,DeFelice:2016cri,Heisenberg:2017hwb,Kase:2017egk,Ramazanoglu:2017xbl,Annulli:2019fzq,Barton:2021wfj,Minamitsuji:2018kof, Herdeiro:2020jzx,Herdeiro:2021lwl,Conlon:2017hhi,Fukuda:2019ewf,dEnterria:2013zqi, Burgess:2020tbq,Heisenberg:1936nmg,Loginov:2015rya,Brihaye:2016pld,Brihaye:2017inn,Herdeiro:2021lwl,Proca:1936fbw,Heisenberg:2014rta, Heisenberg:2016eld, Kimura:2016rzw, Allys:2015sht,Ramazanoglu:2019gbz,Doneva:2022ewd,Unluturk:2023qgk}

When taken at face value, the above problems are sufficient to render self-interacting vectors unphysical, but theories featuring them can also be viewed as effective limits of more fundamental theories, typically called the \emph{UV-complete} (or simply the \emph{UV}) theory. In such a view, the failure of the \emph{effective field theory} (EFT), only an approximation, is not necessarily surprising. Then, a central question for nonlinear extensions of the Proca theory is whether their problems can be explained this way, and if so, how can one obtain a well-posed theory superseding them?

It is already known that arguably the simplest example of self-interacting vector field theories, the so-called \emph{nonlinear Proca theory (NPT)}, is indeed a limiting case of a vector field that gains mass through the Abelian Higgs mechanism for some, but not all, of its parameter space. Hence, the above question seems to be answered in the affirmative in a limited sense. However, this is not the only suggested solution to the problem.

Here, we will address another proposal that aims to continue the time evolution beyond the point where the EFT breaks down: \emph{fixing the equations}, or \emph{fixing} in short~\cite{Israel:1979wp,Barausse:2022rvg,Cayuso:2017iqc,Allwright:2018rut, Cayuso:2020lca,Bezares:2021dma,Franchini:2022ukz,Cayuso:2023aht}. This is a formulaic but ad-hoc method of modifying pathological field equations which are believed to have a well-behaved mother theory, to arrive at some system that, hopefully, better captures the true physics. In some cases, the EFT is known to lack a well-posed time evolution completely, and the fixing provides the said evolution. In others, time evolution is possible but eventually breaks down, and fixing avoids this problem while imitating the EFT closely, we will focus on the latter. The idea goes back to the formulation of relativistic hydrodynamics by~\textcite{Israel:1979wp}, and has more recently been adapted for applications in more general EFTs by~\textcite{Cayuso:2017iqc} 

Fixing has been very successful in mitigating spurious behavior arising from extra degrees of freedom or higher order derivatives~\cite{Allwright:2018rut, Cayuso:2020lca, Franchini:2022ukz}. However, it is not a cure-all and there are, as yet not formalized, conditions for its applicability~\cite{Allwright:2018rut}. Moreover, fixing the non-linear Proca theory, e.g. as suggested in \textcite{Barausse:2022rvg}, goes beyond this. The time evolution problems in this case commonly appears as one approaches the scale at which the EFT becomes strongly coupled, i.e. the scale at which no finite order truncation of the EFT approximates the UV theory, \textit{c.f.} the Planck scale for gravity. Thus, fixing cannot generically capture the dynamics of the UV theory in such cases.\footnote{Fixing is not the sole solution suggested by \textcite{Barausse:2022rvg}, who also suggest careful use of coupling constants and gauges~\cite{Bezares:2020wkn,Bezares:2021dma}. Also see \textcite{Babichev:2018twg}.}

To demonstrate this point, we will compare solution of the NPT (which is an EFT), its fixed versions, and the UV theory (which is known in our case). This will allow us to see some instances where the fixing succeeds, and some where it fails. Surprisingly, in some cases, fixing even performs less well than the unfixed EFT. Our results provide lessons for general applications of fixing beyond the sample cases we present.

\section{The Theories}
We will be studying two sets of theories. The first is a K-essence model, its fixing(s) and UV completion. The second is nonlinear Proca, its fixing and UV completion. In both cases we will be looking at the regimes where a UV completion is known to exist. Thus, we will be able to start with a theory that has no breakdown, and integrate out the heavy degrees of freedom to obtain an EFT that does break down. Finally, we develop the fixed theory by introducing new variables with invented dynamics on top of the EFT, and see how the breakdown can be avoided if we did not know the UV theory. To be exact, the fixed theory can also be considered an EFT, however we will not call it as such to avoid confusion.

This order of presentation is the reverse of the usual course of research. Typically, the EFT is proposed first, and the realization of its problems lead to a search for a UV theory, which cannot be found in many cases unlike our examples. Then, fixing becomes a possible remedy. 

We work in $1+1$ dimensions with $ds^2= -dt^2+dx^2$. $c=1$.

\subsection{Scalar theories}
\label{sec:scalar}
Here we will follow \textcite{Lara:2021piy}, though using some different notation. The UV Lagrangian for the pure scalar set of theories we will be studying is that of a complex scalar field with global $U(1)$ symmetry,
\begin{equation}\label{eq:U1Action}
     \mathcal{L}_{\mathrm{UV}, 1}=-\frac{1}{2} g^{\mu\nu}\partial_\mu\Phib \partial_\nu \Phi -\frac{1}{2} V\left(\Phib \Phi\right)
\end{equation}
with,
\begin{equation}\label{eq:U1 potential}
    V\left(\Phib \Phi\right) = -m^2 \Phib \Phi + \frac{\lambda}{2} \left(\Phib \Phi\right)^2.
\end{equation}
Note $\Phi=0$ is unstable and the global minimum of $V$ occurs at $\Phi = v\exp(i\Theta)$, where $v=\sfrac{m}{\sqrt{\lambda}}$ and $\Theta \in \mathbb{R}$. 

We obtain the so-called \emph{K-essence theory} by integrating out the heavy degree of freedom using the EFT expansion. To do so, write
\begin{equation}
    \Phi = \left(v+\varrho\right)\exp{\left(i \Theta\right)},\quad \Phib = \left(v+\varrho\right)\exp{\left(-i \Theta\right)}.
\end{equation}
In terms of the new variables
\begin{equation*}
    \mathcal{L}_{\mathrm{UV}, 1}= -\frac{1}{2}
    (\partial \varrho)^2 -\frac{1}{2} 
    (v+\varrho)^2 \left[ (\partial \Theta)^2-m^2+ \frac{\lambda}{2} (v+\varrho)^2 \right] ,
\end{equation*}
where $(\partial \varrho)^2 \equiv g^{\mu\nu} \partial_\mu \varrho \partial_\nu \varrho$. This leads to the equation of motion
\begin{equation*}
    \Box\varrho=(v+\varrho) \left[ (\partial \Theta)^2-m^2+ \lambda (v+\varrho)^2 \right]
\end{equation*}
The essence of the EFT expansion is that the dynamics of $\varrho$ is ignorable when it is heavy, i.e. when its effective mass $M \equiv \sqrt{2} m$ is large, holding $\lambda$ fixed (so $v$ diverges with $M$). Quantitatively,
\begin{equation}\label{eq:varrhoEFTKessence}
    (\partial \Theta)^2-m^2+ \lambda (v+\varrho)^2 \approx 0\ \Rightarrow\ \varrho = -\frac{(\partial \Theta)^2}{M^2} +\mathcal{O}\left(\frac{1}{M^4}\right)
\end{equation}
Inserting this back into the Lagrangian, we obtain the EFT
\begin{equation}\label{eq:K-essence action}
    \mathcal{L}_{\mathrm{EFT}, 1}=-\frac{v^2}{2}\left(\partial\Theta\right)^2\left[1-\frac{\left(\partial\Theta\right)^2}{M^2}\right]
\end{equation}
which is the Lagrangian for the so-called \emph{quadratic K-essence}, where the kinetic term is
\begin{equation}\label{eq:KofxDef}
    K(y) = \frac{v^2}{2}\left(y-\frac{y^2}{ M^2}\right)\ \ ,\ \ y =\left(\partial\Theta\right)^2 
\end{equation}
with equation of motion
\begin{equation} \label{eq:K-essence}
   \nabla_\mu\left[ K'(y)\ \partial^\mu\Theta\right]= 0.
\end{equation}
The Cauchy problem here breaks down if $K'\to 0$ and this behavior is what we will seek to fix. It is worth noting at this point that $K'\to 0$ occurs in the regime where $\partial \Theta^2/\left( M^2\right) \sim 1$, \textit{i.e. where the EFT expansion itself breaks down}.

The breakdown can also be seen when we write the equation of motion as
\begin{equation}\label{eq:geff_scalar}
    \gbar^{\mu\nu} \nabla_\mu \nabla_\nu \Theta = 0\ \ ,\ \
    \gbar^{\mu\nu} \equiv g^{\mu\nu} + 2\frac{ K''}{ K'} \nabla^\mu \Theta \nabla^\nu \Theta ,
\end{equation}
where the dynamics is governed by the \emph{effective metric} $\gbar^{\mu\nu}$ rather than the spacetime metric $g^{\mu\nu}$. Hence, time evolution breaks down when $\gbar_{\mu\nu}$ changes signature at $K'=0$.

\textcite{Lara:2021piy} suggested fixing the equation by introducing a new field $\Sigma$ to replace $K'$ in Eq.~\eqref{eq:K-essence}, which is dynamically driven towards $K'$ via
\begin{equation}\label{eq:driver}
    \partial_t\Sigma = -\frac{1}{\tau}\left[\Sigma - K'\right].
\end{equation}
The smaller the timescale $\tau$ the closer this will track the unfixed equations (and the UV completion where applicable). However, $\Sigma=0$ still causes a breakdown of the evolution system, so larger $\tau$ may be required to avoid doing so. In some cases there is no ``good" value of $\tau$, and so we also consider the alternative driver equation
\begin{equation}\label{eq:alternativedriver}
    \partial_t\Sigma = -\frac{1}{\tau}\left(\frac{2 K'}{v^2}\right)^2\left[\Sigma - K'\right] .
\end{equation}

\subsection{Vector theories}
\label{sec:vector}
This proceeds almost identically to the previous case, starting from the gauged version of the Lagrangian in Eq.~\eqref{eq:U1Action}. In other words, the UV theory is the Abelian Higgs model for a vector field $A^\mu$,
\begin{align}\label{eq:HiggsAction}
    \mathcal{L}_{\mathrm{UV}, 2}= -\frac{1}{4} \left[ F_{\mu\nu}F^{\mu\nu}
    +2\overline{D_\mu\Phi} D^\mu\Phi +2 V\left(\Phib \Phi\right) \right]
\end{align}
where $F_{\mu\nu} = \nabla_\mu A_\nu -\nabla_\nu A_\mu$ and
\begin{equation}
    D_\mu\Phi = \partial_\mu\Phi - i q A_\mu \Phi .
\end{equation}
This guarantees invariance under the $U(1)$ transformation,
\begin{equation}
    A_\mu\to A_\mu +\partial_\mu f,\quad \Phi \to \Phi \exp{\left(i q f\right)}.
\end{equation}
It is also useful to define the gauge invariant field
\begin{equation}
    X_\mu \equiv A_\mu - \frac{1}{q}\partial_\mu\Theta ,
\end{equation}
which will take the place of ($-1/q$ times) $\partial_\mu\Theta$ from the scalar section. Indeed, writing $\Phi=\left(v+\varrho\right)\exp{\left(i \Theta\right)}$, and performing the same expansion to integrate out $\varrho$, the EFT Lagrangian becomes
\begin{equation}\label{eq:EFT_action}
     \mathcal{L}_{\mathrm{EFT}, 2} = -\frac{1}{4} F_{\mu\nu}F^{\mu\nu} - K(q^2 X^2) ,
\end{equation}
with $K$ the same as in Eq.~\eqref{eq:KofxDef} and $X^2=X_\mu X^\mu$. This is exactly NPT with  the squared mass and the quartic coupling
\begin{equation}\label{eq:EFT_eom}
    \mu_\mathrm{NPT}^{\phantom{\mathrm{NPT}}2}=q^2v^2,\quad \lambda_\mathrm{NPT}=-\frac{2q^2}{M^2},
\end{equation}
respectively, recalling that the coefficient of the quartic term is $\lambda_\mathrm{NPT} \mu_\mathrm{NPT}^{\phantom{\mathrm{NPT}}2}/4$ in that convention~\cite{Clough:2022ygm,Coates:2022qia}.

Note that the UV completion only exists for $\mu_\mathrm{NPT}^{\phantom{\mathrm{NPT}}2}>0$ and $\lambda_\mathrm{NPT}<0$, which is only a part of the parameter space. The full equations of motion are,
\begin{equation}
    \nabla_\mu F^{\mu\nu}=2q^2K' X^\nu,\quad \nabla_\mu\left(K'X^\mu\right)=0,
\end{equation}
which, again, will break down where $K'=0$, this time the effective metric being~\cite{Coates:2022qia,Coates:2022nif}
\begin{equation}
    \gbar^{\mu\nu}= (1+\lambda_\mathrm{NPT} X^2) g^{\mu\nu} + 2\lambda_\mathrm{NPT} X^\mu X^\nu .
\end{equation}

Therefore we will again use some fixing, one suggested by \textcite{Barausse:2022rvg}. We split the vector field via the Stueckelberg mechanism, $X_\mu \equiv A_\mu - (1/q)\partial_\mu\Theta$, and introduce an additional field $\Sigma$ to take the place of $K'$
\begin{align}
    \nabla_\mu F^{\mu\nu} &=2q^2\Sigma X^\nu,\quad \nabla_\mu\left(\Sigma X^\mu\right)=0 \nonumber \\
    \partial_t\Sigma &= -\frac{1}{\tau}\left[\Sigma-K'\right],\quad X_\mu \equiv A_\mu - \frac{1}{q}\partial_\mu\Theta .
    \label{eq:fixed_eom}
\end{align}

\section{Results}
\label{sec:results}
For various scenarios, we will compare the time evolution (at the sample fixed point $x=1$) for the EFT, the fixed theory (for various $\tau$) and the UV theory for both the scalar and vector fields, starting from the same initial data. For example, the standard expectation is that the vector field $X_\mu$ of NPT and the gauge invariant vector fields $X_\mu=A_\mu - (1/q) \partial_\mu\Theta$ for the fixed and UV theories agree with each other when the approximations work as they are supposed to. We also track the determinant of the effective metric, $\gbar$, whose vanishing implies the breakdown of time evolution for the EFT. Our results seem quite general based on our explorations. Numerical errors are smaller than the thickness of the lines in the following plots,\footnote{Exceptions to this such as cases where the computation breaks down will be apparent in our discussion.} hence, we do not provide explicit error bars. Details of the numerics~\cite{Zilhao:2015tya, Clough:2022ygm,Coates:2022qia} and the solutions are in the appendices.

The first result is that the right fixing can do a very good job of recreating the behavior of the UV theory after the EFT breaks down, but in all cases we explored, this only happens if the region of interest ``settles down'' \emph{quickly}. As we are working in $1+1$ dimensional Minkowski space, settling down means when the fields disperse and head off to infinity. When there is non-trivial behavior considerably beyond when the breakdown occurs for the EFT, the fixing eventually fails to capture the UV theory due to the excitation of high frequency modes that fixing cannot replicate. Indeed, it seems there is an inevitable excitation of some high frequency mode in the UV when the EFT breaks down, which also quickly causes the numerics to fail as well. Success of fixing observed by others in analogous cases, e.g. \textcite{Lara:2021piy}, is likely related to the formation of a horizon in their work, which can keep the high frequency behavior from influencing the exterior solution.

\begin{figure}
\begin{center}
\includegraphics[width=0.48\textwidth]{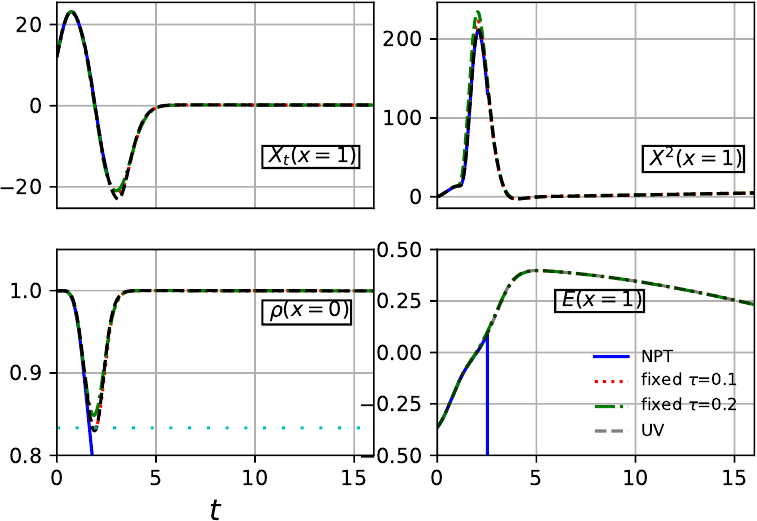}
\end{center}
\caption{Collision of two pulses in the vector field theory where the EFT barely breaks down at $t\approx 2.5$, but the fields disperse afterwards in the UV theory. Fixing works successfully as expected. The horizontal dotted line is the $\varrho$ value where the breakdown is theoretically expected.} 
\label{fig:colliding_pulses}
\end{figure}
An example of fixing working well at extending the solution beyond the breakdown of the EFT is in Fig.~\ref{fig:colliding_pulses}. Here, two Gaussian vector field pulses collide, and the resulting high field amplitudes when they meet barely break down the EFT. The fixed theory continues to evolve while tracking the UV theory closely. This follows our findings, since the high amplitudes responsible for the breakdown occur for a short duration in the UV-theory, and the fields quickly settle down after the pulses pass through each other.

\begin{figure}
\begin{center}
\includegraphics[width=0.48\textwidth]{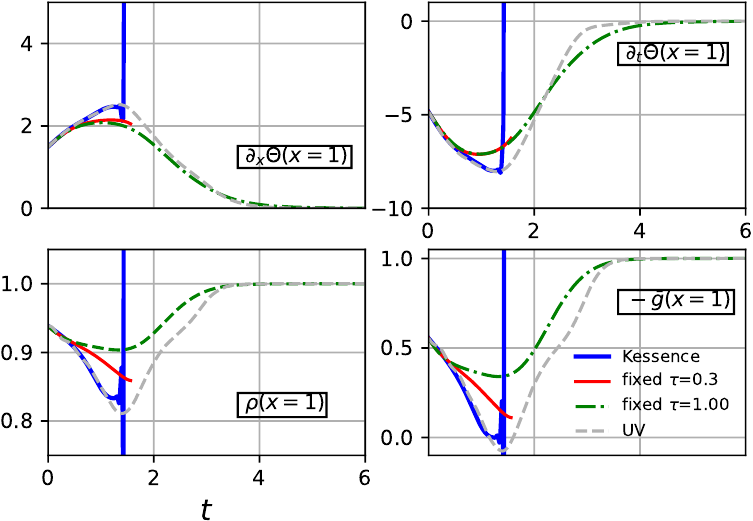}
\end{center}
\caption{Sample evolution of the scalar theory where fixing does not work well for any value of $\tau$, due to slow settling down. K-essence breaks down at $t \approx 1.5$, the fixed versions [Eq.~\eqref{eq:driver}] either also break down (in red) or else do not approximate the UV solution well (green).}
\label{fig:standard_fix}
\end{figure}

Fig.~\ref{fig:standard_fix} is a case where the fixing fails. Here, the fields settle down, but not fast enough, in a scenario where a high value of $\dot{\Theta}$ pushes the initial data towards breakdown. The ``fixed" evolution still breaks down for small $\tau$. If we use larger $\tau$ to avoid this, the UV theory is not well approximated except at late times when all fields return to vacuum. We would be unaware of the latter failure if we did not already know the UV theory. In some instances, the fixed theory fails even when the EFT does not, where numerical dissipation~\cite{Kreiss73,shibata2015numerical} seems to play an important role, see Appendix.~\ref{sec:dissipation_fixing}.

\begin{figure}
\begin{center}
\includegraphics[width=0.48\textwidth]{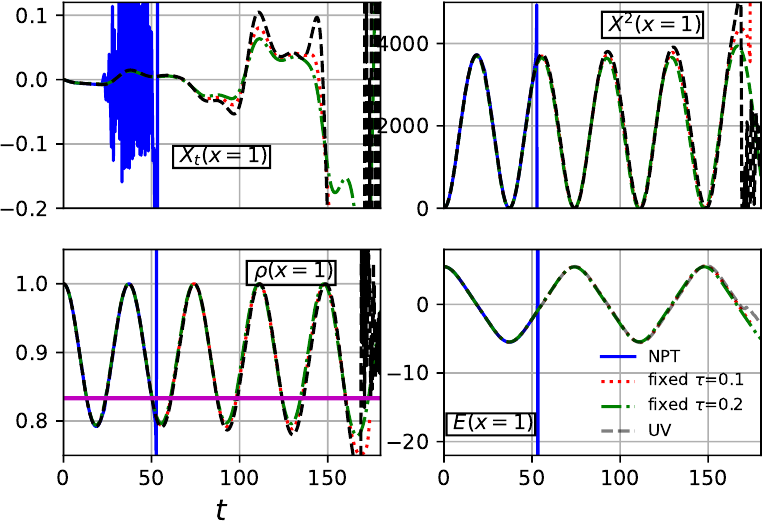}
\end{center}
\caption{Continuous oscillations induced by a constant background electric field, leading to the vector field theory not settling down promptly. The EFT breaks down within the first cycle. After a few cycles, some high frequency component dominates the UV theory solution, at which point the fixed theory also breaks down (noisy pattern after $t \approx 170$). For all initial data of this form, either the dip into the breakdown regime is too mild to break the EFT, or else this high frequency component eventually comes to dominate the UV solution, which later crashes our numerics.}
\label{fig:high_frequency}
\end{figure}
Fig.~\ref{fig:high_frequency} shows a case which does not settle down to vacuum, where there are sustained oscillations due to an initial constant background contribution to the electric field. This time, fixing fails due to a flow of energy to high frequencies. This is not necessarily surprising since the UV theory is nonlinear, however, the fact that not settling to vacuum is correlated with such behavior, which guarantees the failure of fixing, is noteworthy.\footnote{Once the high frequency dominates, the numerical results are not reliable any further. However, the energy flow to high frequencies is already apparent before this happens, where the numerics converge without issue. See the appendix for details.}

\begin{figure}
\begin{center}
\includegraphics[width=0.48\textwidth]{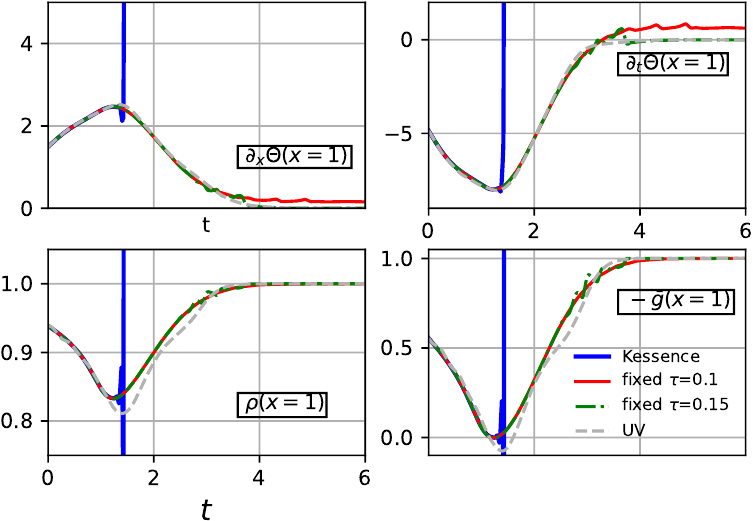}
\end{center}
\caption{The same as in Fig.~\ref{fig:standard_fix}, but with the alternative driver Eq.~\eqref{eq:alternativedriver}. This is a better fixing than that of Fig.~\ref{fig:standard_fix}, since the fixed and UV theories agree more closely.}
\label{fig:alt_fix}
\end{figure}
The second main result is that the dynamics of the fixed case depends strongly on the choice of the parameter $\tau$ and the particular driver equation used. The best value of $\tau$, i.e. the smallest that does not lead to a breakdown due to $\Sigma = 0$, for different initial data  can range over several orders of magnitude. In some situations there is no particularly good choice of $\tau$ that we could find. That is, there is no value of $\tau$ which both avoids $\Sigma = 0$ and closely tracks the UV solution, as we saw in Fig.~\ref{fig:standard_fix}. However, we did find that using an alternative driver equation can sometimes help as in Fig.~\ref{fig:alt_fix}. Without knowing the UV completion, there is no benchmark against which to compare the results of the fixed equations in general. This raises a seemingly unanswerable questions as to whether fixed solutions are physically realistic generically. We have seen in Fig.~\ref{fig:standard_fix} that the fixing~\eqref{eq:driver} is not, in this specific case.

\section{Discussion}
We have seen that fixing the equations approach is not universally reliable.  In certain situations, the simplest application of fixing can lead to scenarios where no good fixing parameter exists. Therefore without some mathematical theory of fixing, it is not a dependable tool for generating physically relevant solutions. In fact, even given some such understanding, it does not seem to be fit for purpose in all cases. That is, if one wishes to find an efficient method for generating EFT solutions, fixing can be quite ineffective. This is because, to solve an EFT to $m$-th order using the perturbative approach, which we have good reason to believe converges whenever EFTs are appropriate \cite{Reall:2021ebq}, one needs to solve $p$ systems of $N$ equations. Whereas to apply the fixing method one needs to solve $N+p$ equations to however many times it takes to solve an optimization problem in $s$ continuous variables. The use case is then whenever $p\gg s$, and how often that can actually be achieved will not be known until a theory of fixing has been developed beyond the current heuristic approach. 

We reiterate that our findings are not in contradiction with the existing specific results obtained via fixing, where some general guidelines have been shown to be useful in specific cases~\cite{Bezares:2021dma,Bezares:2021yek}. Rather, we show that the technique can not be assumed to be generically useful. We considered $1+1$ dimensions here, but loss of hyperbolicity studies indicate that the behavior is similar in any dimension~\cite{Coates:2022qia}.

A feature we have seen in any solution that does not quickly settle to vacuum is that the breakdown of the EFT in our cases always coincided with the excitation of a high frequency mode in the UV theory that eventually overwhelmed our numerics. Although we cannot see an analytic reason for this, it is perhaps no surprise that the breakdown of the EFT coincides with the inevitable excitation of modes that have been integrated out. We believe this is another mark against fixing the equations as the technique cannot pick up on this by definition. Without the UV solution to compare with, it does not seem possible to know when this suppressed behavior will become important, beyond simple order estimates of cutoff scales (see Appendix.~\ref{sec:numerics_tests} for some more detail). It finally is worth mentioning that the numerical threshold for this behavior seemed to be equal to the numerical threshold for the breakdown of the EFT. In other words, if the solution barely dipped into the breakdown regime for a small enough time such that the EFT continued to evolve afterwards, then we could also not see this high frequency behavior in the UV theory, though this behavior may not be universal~\cite{Bezares:2020wkn}.

\acknowledgements
We thank Will East for many valuable discussions and stimulating questions, and Marco Crisostomi for his comments on the manuscript. F.M.R acknowledges support from T\"UB\.ITAK Project No. 122F097.

\appendix

\section{First order formulation of the time evolution}

In all cases, we use a $1+1$ decomposition of the flat background spacetime using the trivial foliation, that is the normal vector to the spatial slices is simply $n^\mu = (\partial_t)^\mu$, which greatly simplifies the equations of motion. See \textcite{Coates:2022qia} or \textcite{Clough:2022ygm} for details of the foliation procedure for a general curved spacetime. We will use an overdot to symbolize time derivative: $\dot{\pi} \equiv \partial_t \pi$.

Our time evolution formulation is identical to that of \textcite{Coates:2022qia} for the case of the nonlinear Proca theory, i.e. the EFT. The standard $1+1$ decomposition for Eqs.~\eqref{eq:EFT_action} and~\eqref{eq:EFT_eom} uses the variables
\begin{align}
 X_\mu = \left(-\varphi_X, \cala_X \right),\quad F_{tx} = - E_X
\end{align}
and leads to the field equations~\cite{Coates:2022qia,Clough:2022ygm}
\begin{subequations}
\begin{align}
     m^2 \bar{g}_{tt} \dot{\varphi}_X &=  - m^2 \bar{g}_{tt}\partial_x \cala_X+4q^2 \cala_X \varphi_X \partial_x \varphi_X \nonumber \\
     &+2 q \cala_X E_X \varphi_X - \zeta_X\\
    \dot{\cala}_X &= - E_X - \partial_x\varphi_X\\
    \dot{E}_X &= \frac{q^2  m^2 z_X}{\lambda}\cala_X +\partial_x\zeta_X\\
    \dot{\zeta}_X &= -\kappa\zeta_X +\mathcal{C}_X \\ \mathcal{C}_X &= \partial_x E_X +\frac{q^2 m^2 \varphi_X z_X}{\lambda} 
\end{align}
\end{subequations}
where

\begin{align}
   m^2 \bar{g}_{tt}&= (-m^2 + q^2(\cala_X^2-3\varphi_X^2)),\\
   z_X &=1 +\frac{q^2}{m^2}(\varphi_X^2-\cala_X^2)
\end{align}
${\cal C}_X$ is the constraint, and $\zeta_X$ is the constraint damping parameter~\cite{Coates:2022qia,Clough:2022ygm}.

We perform a similar decomposition for the the UV Abelian Higgs theory of Eq.~\eqref{eq:HiggsAction}, and also separate the scalar $\Phi=\Phi_r+i\Phi_i$ into its real and imaginary parts
\begin{align}
    A_\mu = \left(-\varphi, \cala \right),\ F_{tx} = - E,\ \dot{\Phi}_r = Q,\ \dot{\Phi}_i = J\ .
\end{align}
Unlike the EFT, we need a gauge choice for the Higgs theory. For this purpose, we impose the Lorenz gauge, $\nabla_\mu A^\mu =0$, which provides the time evolution system
\begin{subequations}
\begin{align}
    \dot{\varphi} &= - \partial_x \cala\\
    \dot{\cala} &= - E - \partial_x\varphi\\
    \dot{E} &= q^2\cala\rho^2+q\left(\Phi_i\partial_x\Phi_r -\Phi_r\partial_x\Phi_i\right)+\partial_x\zeta\\
    \dot{\Phi}_r &= Q\\
    \dot{\Phi}_i &= J\\
    \dot{Q} &= +2 q \varphi J  + \left[m^2z-\lambda \Phi_i^2\right]\Phi_r -\lambda \Phi_r^3 \nonumber\\
    &\hphantom{=}\,\,+2 q\cala\partial_x\Phi_i +\partial_x^2\Phi_r\\
    \dot{J} &= -2 q \varphi  Q  + \left[m^2z-\lambda \Phi_r^2\right]\Phi_i -\lambda \Phi_i^3\nonumber\\
     &\hphantom{=}\,\,-2 q\cala\partial_x\Phi_r +\partial_x^2\Phi_i\\  
    \dot{\zeta} &= -\kappa\zeta +\mathcal{C}\\ 
    \mathcal{C} &= \partial_x E +q^2\varphi\left(\Phi_r^2+\Phi_i^2\right)+q \left(\Phi_r J - \Phi_i Q\right).
\end{align}
\end{subequations}
In this case,
\begin{align}\label{eq:higgs_z}
   z&=1 +\frac{q^2}{m^2}(X_t^2-X_x^2)\ .
\end{align}
Recall that
\begin{equation}\label{eq:higgs_X}
    X_\mu \equiv A_\mu - \frac{1}{q}\partial_\mu\Theta
\end{equation}
is the gauge invariant vector field, where we also define the derived variables
\begin{align}
\rho = \sqrt{\Phi_r^2 +\Phi_i^2}, \quad\Theta = \tan^{-1}\left(\frac{\Phi_i}{\Phi_r}\right)
\end{align}
which are important to compare the three theories with each other.

The variables for the fixed theory of Eq.~\eqref{eq:fixed_eom} are similar to the Higgs theory
\begin{align}
    A_\mu = \left(-\varphi_f, \cala_f \right), \quad F_{tx} = - E_f
\end{align}
with the differences that $\Sigma$ is introduced and evolved with its own ad-hoc time evolution, and we have $\Theta$ as an independent variable, whose dynamics is determined by the Lorenz condition $\nabla_\mu\left(\Sigma X^\mu\right)=0$. The equations of motion are
\begin{subequations}
\begin{align}
    \dot{\varphi}_f &= - \partial_x \cala_f \\
    \dot{\cala}_f &= - E_f - \partial_x\varphi_f \\
    \dot{E}_f &= 2 q^2 \Sigma X_x +\partial_x\zeta_f\\
    \dot{\Theta} &= \gamma\\
    \dot{\gamma} &= \frac{-q X_t}{\tau}\left(1 -\frac{m^2 z}{2\lambda\Sigma}\right)-\frac{q X_x\partial_x\Sigma-\Sigma \partial_x^2\Theta }{\Sigma}\\ 
    \dot{\Sigma} &= -\frac{1}{\tau}\left[\Sigma -\frac{m^2 z}{2\lambda} \right]  \\
    \dot{\zeta}_f &= -\kappa\zeta +\mathcal{C}, \quad \mathcal{C}_f = \partial_x E_f -2 q^2 \Sigma X_t
\end{align}
\end{subequations}
Here, we abused the notation, and used $X_t$, $X_x$ and $z$ as auxiliary variables, even though they were already defined for the Higgs theory. This is because  Eqs.~\eqref{eq:higgs_z} and~\eqref{eq:higgs_X} are valid exactly in the same form for the fixed equations as well, with the difference that $\Theta$ is one of our primary variables in this case.

To compare the three theories, we use
\begin{align}
    X_{\mathrm{EFT},\mu} &= \left(-\varphi_X,\ \cala_X \right) \\
    X_{\mathrm{UV},\mu} &= \bigg(-\varphi-\frac{1}{q}\partial_t \tan^{-1}\left(\frac{\Phi_i}{\Phi_r}\right), \nonumber \\ & \phantom{= \bigg(} \cala-\frac{1}{q}\partial_x \tan^{-1}\left(\frac{\Phi_i}{\Phi_r}\right) \bigg) \\
    X_{\mathrm{fixed},\mu} &= \left(-\varphi_f-\frac{1}{q}\partial_t \Theta,\ \cala_f-\frac{1}{q}\partial_x \Theta \right)
\end{align}

Recalling that the vector theory is the gauged version of the scalar one, the scalar evolution equations are obtained by setting all vector related variables to zero. That is, we simply set $q=\varphi=\cala=E=0$ in the UV theory, and likewise in the fixed theory and the EFT.

\section{Initial data}
\label{sec:initial_data}

The initial data is chosen so that all fields (that exist in every theory) match,
\begin{align}
    &\Theta_\mathrm{UV}(0) = \Theta_\mathrm{EFT}(0) =\Theta_\mathrm{fixed}(0)\\
    &X_\mathrm{UV}(0) = X_\mathrm{EFT}(0) =X_\mathrm{fixed}(0),
\end{align}
and are the same for the relevant time derivatives (in the guise of $\dot{\Theta}$ and the electric field $E$).

The initial value for the UV field and its time derivative will be given by,
\begin{align}  
    \rho(0) = (v+\varrho(0)) =& v\sqrt{1+2\varrho_\mathrm{EFT}(0)/v}  \nonumber\\
    = & v + \varrho_\mathrm{EFT}(0) +\mathcal{O}\left(\sfrac{1} {M^2}\right),\label{eq:rho_0} \\
    \dot{\rho}(0)= &\dot{\varrho}_\mathrm{EFT}(0)\label{eq:rhodot_0}
\end{align}
where $\varrho_\mathrm{EFT}$ is given by Eq.~\eqref{eq:varrhoEFTKessence} or 
\begin{equation}\label{eq:varrhoEFTProca}
    \varrho = -q^2 \frac{X_\mu X^\mu}{M^2} +\mathcal{O}\left(\frac{1}{M^4}\right)
\end{equation}
for the vector field, as appropriate. This specific form is chosen to satisfy the constraint equation in the nonlinear Proca case. Finally, for the fixed equations we pick,
\begin{equation}
    \Sigma(0)=K'\left(y(t=0)\right),
\end{equation}
with $y=(\partial\Theta)^2$ or $X^2$ as appropriate. 

To create Figs.~\ref{fig:standard_fix},~\ref{fig:alt_fix} and~\ref{fig:dissipation_fix}  we choose initial data for K-essence of the form,
\begin{equation}
    \Theta = C_\Theta\ \mathrm{erf} \left(\frac{x}{\sqrt{2}\sigma}\right), \quad\dot{\Theta} = C_\gamma \exp{-\frac{x^2}{2\sigma^2}}
\end{equation}
with $\sigma = 1$, $C_\gamma=-8$, $C_\Theta \approx 12.4$\footnote{$C_\Theta = \sigma\sqrt{\frac{\pi}{6}\left(3C_\gamma^2 + m^2\right)}-0.0001$}, in the fixed theory we additionally initialize $\Sigma=K'$, and for the UV theory we initialize $\Phi_r$ and $\Phi_i$ so that 
\begin{equation}
    \Phi = \rho_\mathrm{EFT}\exp{i\Theta},\quad\dot{\Phi} = \left(\dot{\rho}_\mathrm{EFT}+i\rho_\mathrm{EFT}\dot{\Theta}\right)\exp{i\Theta}\nonumber
\end{equation}

The initial data for Figs.~\ref{fig:colliding_pulses} and~\ref{fig:colliding_pulses_perturbed} are the same for the non-linear Proca theory and its fixed versions,
\begin{equation}
    \varphi=-C_\cala x^3 \exp{-\frac{x^2}{2\sigma^2}},\quad \cala = C_\cala |x|^3 \exp{-\frac{x^2}{2\sigma^2}}
\end{equation}
with $C_\cala = 20$, $\sigma = 1$. The electric field $E$ is calculated analytically from the constraint with the choice $E(x) \to 0$ as $x\to\infty$. For the fixed versions we again initialize $\Sigma=K'$. We choose initial data for the UV theory such that $\Phi_i=0$, $\dot{\Phi}_i=0$, $\Phi_r = \rho(0)$ and $\Phi_r = \dot{\rho}(0)$ as given by Eqs.\eqref{eq:rho_0} and~\eqref{eq:rhodot_0}. For Fig.~\ref{fig:colliding_pulses_perturbed} we perform the modification
\begin{equation}
    \dot{\rho}(0)\to 1.5\ \dot{\rho}(0) + 0.05.
\end{equation}

Fig.~\ref{fig:high_frequency} has initial data for the non-linear Proca theory of the form
\begin{equation}
    \varphi=C_\varphi x \exp{-\frac{x^2}{2\sigma^2}},\quad \cala = C_\cala  \exp{-\frac{x^2}{2\sigma^2}}
\end{equation}
with $\sigma = 10$, $C_\varphi = 0$, $C_\cala = 0.1 $ and the electric field $E$ again given by the analytic solution to the constraint equation. This time however we shift the electric field by a constant such that $E\to 5.5$ as $x\to\infty$.

For all numerical results in the figures, we use $m=10,$ $\lambda=100$ and $q=0.1$.

\section{Numerical implementation and tests} 
\label{sec:numerics_tests}
%
\begin{figure}
\begin{center}
\includegraphics[width=0.48\textwidth]{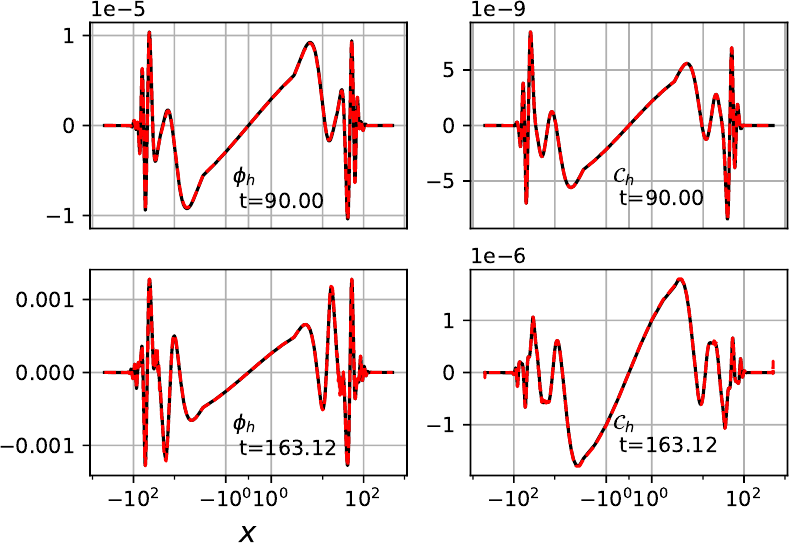}
\end{center}
\caption{Numerical convergence analysis of Fig.~\ref{fig:high_frequency} for the UV theory using the truncation error estimates $\phi_{2h}-\phi_{h}$ and the constraints ${\cal C}_h$, where a subscript $h$ denotes the numerical result when a  spatial step size $h$ is used. There is convergence throughout as shown by snapshots in the middle of the computation (upper row), and near the breakdown (lower row). Left column: $16(\phi_{2h}-\phi_h)$ (red) and $\phi_{4h}-\phi_{2h}$ (black), whose agreement demonstrates fourth order convergence as expected ($h=2^{-3}$.). Right column: $16 {\cal C}_{2h}$ and ${\cal C}_h$ agree, since the constraint itself should converge to zero. The convergence is broken around $t \approx 170$ due to growing high frequency features (see Fig.~\ref{fig:problem_high_frequency} and the related discussion).
}
\label{fig:higgs_convergence}
\end{figure}
\begin{figure}
\begin{center}
\includegraphics[width=0.48\textwidth]{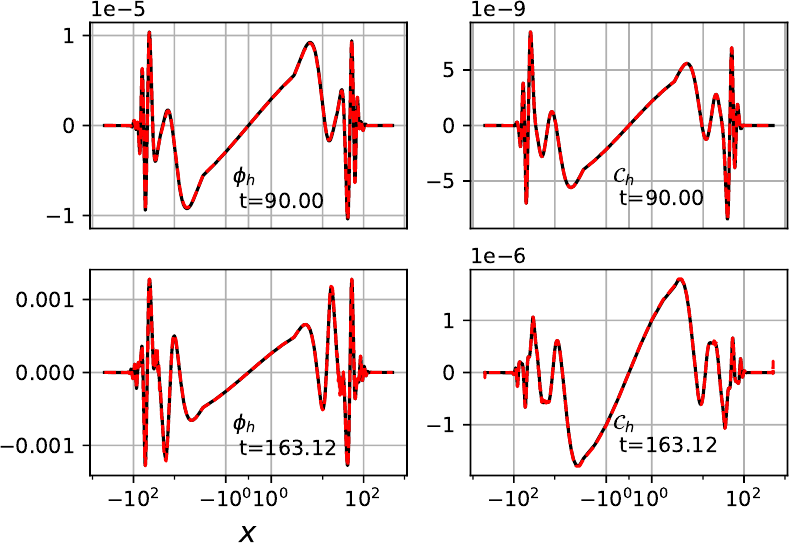}
\end{center}
\caption{Same as Fig.~\ref{fig:higgs_convergence} but for the fixed theory with $\tau=0.1$.}
\label{fig:fixed_convergence}
\end{figure}
We follow the numerical methods of \textcite{Coates:2022qia}. Namely, we use the method of lines with fourth order Runge-Kutta integration for time evolution and fourth order stencils for spatial derivatives. We use constraint damping with $\kappa=1$ to control the constraint~\cite{Zilhao:2015tya, Clough:2022ygm}. We also use Kreiss-Oliger dissipation~\cite{Kreiss73} to avoid numerical high frequency noise. Figs.~\ref{fig:standard_fix} and~\ref{fig:alt_fix} use $\epsilon= 2^6\sigma_{\rm KO}=0.1$~\cite{shibata2015numerical}, Fig.~\ref{fig:colliding_pulses} uses $\epsilon=0.3$, and Fig.~\ref{fig:high_frequency} uses $\epsilon=0.5$. 

We test and confirm the convergence of our solutions by repeating the computation at different resolutions. You can see sample cases in Figs.~\ref{fig:higgs_convergence} and~\ref{fig:fixed_convergence}. We observe the expected fourth order convergence up to the breakdown of time evolution in the EFT and fixed theories. The difference between numerical solutions at different resolutions provides an estimate of the the numerical truncation error, which is too small to show in the figures of the main text as previously mentioned. The UV theories have indefinite time evolution in principle, but as we discussed in relation to Fig.~\ref{fig:high_frequency}, in certain cases a high frequency feature develops in late times, which causes our numerics to crash as well, which we will discuss further shortly.

\begin{figure}
\begin{center}
\includegraphics[width=0.48\textwidth]{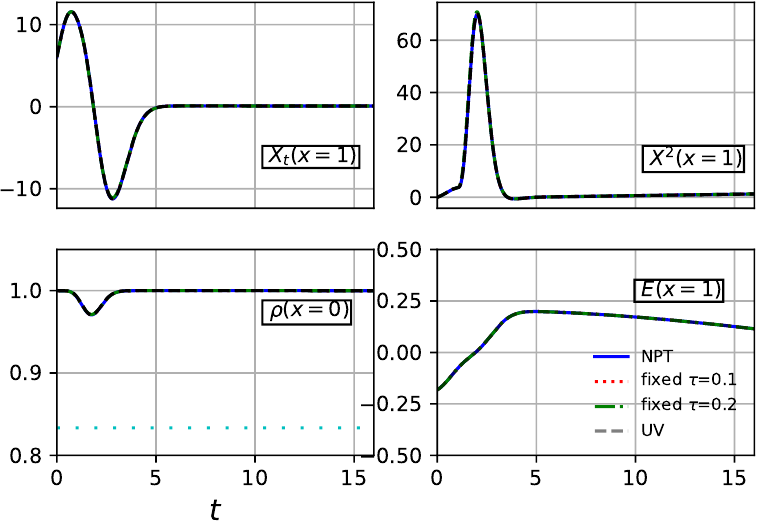}
\end{center}
\caption{Demonstration of the agreement of all three theories when there is no breakdown and the heavy mode is essentially non-dynamical. This is a collision of two vector field pulses as in Fig.~\ref{fig:colliding_pulses}, but the initial amplitudes are lower so that there is no breakdown. The horizontal dotted line is the $\varrho$ value where the breakdown is theoretically expected.}
\label{fig:everyhing_working}
\end{figure}
Aside from the numerical tests, we also compared the EFT, the fixed theory and the UV theory with each other for both scalar and vector fields, and for various cases. When there is no breakdown, we expect these to provide the same results for gauge-independent quantities. In the extreme limit of applicability of the EFT, i.e. when the heavy degree of freedom is almost non-dynamical and fixed at its equilibrium value, this agreement can be seen directly in Fig.~\ref{fig:everyhing_working}. 

\begin{figure}
\begin{center}
\includegraphics[width=0.48\textwidth]{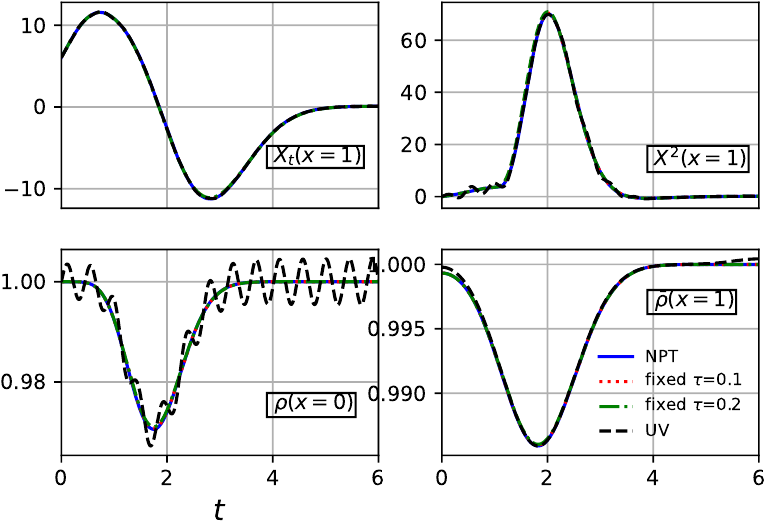}
\end{center}
\caption{Sample evolution of the vector theory where the initial conditions for the UV theory are perturbed, causing oscillatory behavior of $\rho$ around the EFT solutions, in a regime far from breakdown. EFT captures the coarse-grained behavior of the heavy field as expected. This can be seen either visually in the lower left or explicitly in the lower right, where we show the coarse-grained $\varrho$, denoted by $\bar{\varrho}$, obtained by convolution with a window function. Deviations near the beginning and end of the evolution are due to coarse-graining.}
\label{fig:colliding_pulses_perturbed}
\end{figure}
The agreement of the theories can still be observed when the heavy degree of freedom has noticeable oscillations around its equilibrium, however, in this case, one needs to average out the UV theory fields in a coarse-graining procedure for comparison, as in Fig.~\ref{fig:colliding_pulses_perturbed}. We performed the coarse-graining by convolving $\rho(t,x)$ with a two-dimensional Gaussian filter, with standard deviations of $\sigma_x=\sigma_t=0.5$ in the $x$ and $t$ directions, respectively. These values are chosen such that the window averages the local dynamics over a few cycles of the oscillating $\rho$, but it is also not overly wide to average out the whole global evolution.

\begin{figure}
\begin{center}
\includegraphics[width=0.48\textwidth]{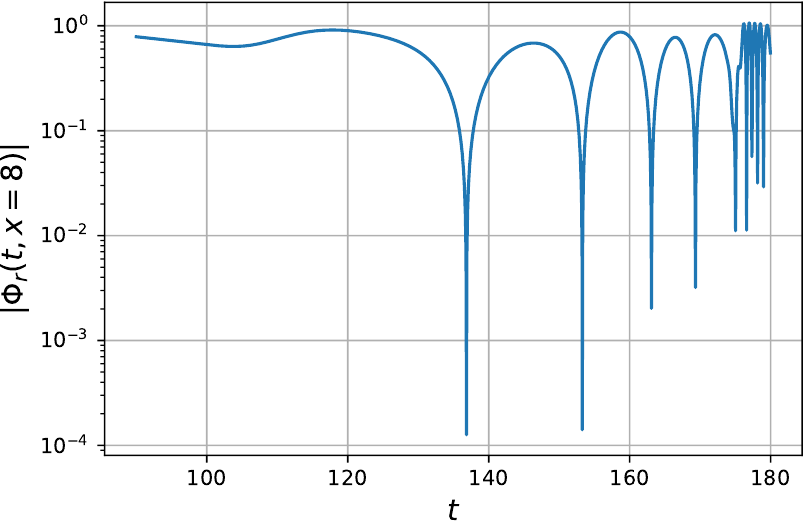}
\end{center}
\caption{Time evolution of $|\Phi_r|$ at $x=8$ for the UV theory in Fig.~\ref{fig:high_frequency}. This clearly shows the growth of a high frequency mode, which eventually causes the numerical computation to crash.}
\label{fig:problem_high_frequency}
\end{figure}

Finally we demonstrate that the behavior in Fig.~\ref{fig:high_frequency} is really due to a transfer of energy to high frequencies. By focusing on, say, the real part of the scalar field $\Phi_r$ we can see this directly in Fig.~\ref{fig:problem_high_frequency}. Note that even though the high frequency component starts to be visible earlier in Fig.~\ref{fig:problem_high_frequency}, its effect on numerics becomes noticeable only later when the computation crashes, e.g. the numerical solutions still converge as expected at $t \approx 165$ (see Fig.~\ref{fig:higgs_convergence}).
%

\section{Numerical dissipation as fixing}
\label{sec:dissipation_fixing}
%
\begin{figure}
\begin{center}
\includegraphics[width=0.48\textwidth]{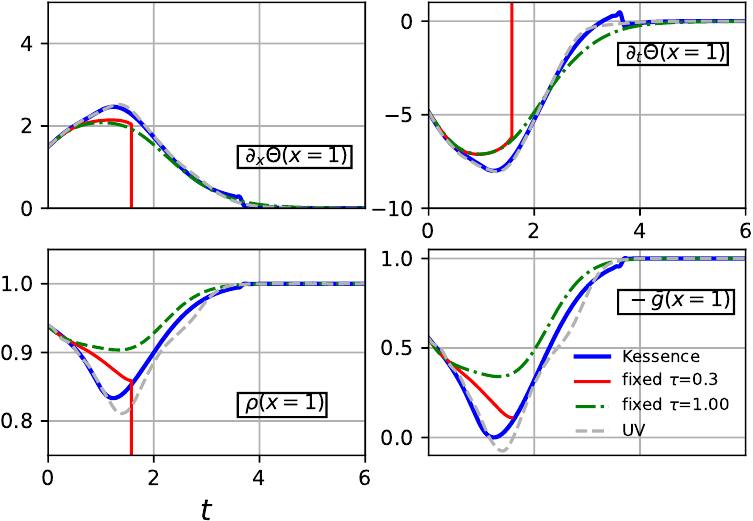}
\end{center}
\caption{Same as Fig.~\ref{fig:standard_fix}, but with increased numerical dissipation, which acts as fixing on the EFT. Recall that Fig.~\ref{fig:standard_fix} is an example of scalar theory evolution where fixing does not work well for any value of $\tau$. However, increased numerical dissipation here can act as fixing and prevent the EFT (K-essence in this case) from breaking down while closely tracking the UV theory.
}
\label{fig:dissipation_fix}
\end{figure}
Finally, we would like to report an observation on the interplay of numerical dissipation and fixing. Fixing typically avoids the ill-posed behavior of an EFT by controlling the growth of high frequency modes. In this sense, its essence is similar to numerical dissipation, where an artificial viscosity term is added to the numerical implementation of the field equations to avoid high frequency numerical errors~\cite{Kreiss73,shibata2015numerical}. 

A successful numerical dissipation aims to change the equation at the level of truncation error, but it can also potentially suppress physically existing high frequency modes. This latter effect is generally unwanted, however, it can act as a form of fixing, depending on how much artificial viscosity is used. Fig.~\ref{fig:dissipation_fix} shows an explicit example. This is the same initial data as in Fig.~\ref{fig:standard_fix}, but the dissipation parameter $\epsilon= 2^6\sigma_{\rm KO}$~\cite{shibata2015numerical} is increased from $0.1$ to $0.3$. Recall that fixing does not work very well for any $\tau$, in the sense that it does not closely follow the UV-complete evolution in Fig.~\ref{fig:dissipation_fix}. However, the unfixed EFT (K-essence), whose time evolution breaks down for low dissipation in Fig.~\ref{fig:standard_fix}, can now evolve without issue, while agreeing with the UV theory quite well. Hence, increased dissipation has effectively fixed K-essence without the need to prescribe an ad-hoc dynamics for $\Sigma$.

Fixing-by-dissipation is likely only meaningful in examples where the EFT is ``mildly'' broken, in the sense that the UV theory spends a very short amount of time in the problematic region (where the determinant of the effective metric $\gbar$ changes sign in our example). This is because numerical dissipation contributes perturbatively to the time evolution since it is supposed to act only at the level of the truncation error. Hence, the difference in the solution by increasing the dissipation is small, which can only help avoiding the breakdown if $\gbar$ becomes positive for a short time and satisfying $\gbar \ll 1$. This is not necessarily surprising, since such time evolution problems are a prime target of fixing schemes in general.

We have investigated a single case here, but it remains to be seen if a fixing-by-dissipation scheme can be formalized for general purposes.

\bibliography{references_all}
\end{document}